\documentclass[conference,10pt]{IEEEtran}
\usepackage{color}
\usepackage{graphicx}
\usepackage{amsmath}
\usepackage{amsthm}
\usepackage{amssymb}
\usepackage{amsfonts}
\usepackage{mathtools}
\usepackage{listings}
\usepackage{flushend}
\usepackage{fancybox}
\usepackage{cite}
\usepackage{tikz}
\usetikzlibrary{arrows,matrix,positioning}

\DeclareMathOperator{\vecc}{vec}
\renewcommand{\vec}[1]{\ensuremath{\mathbf{#1}}}

\newcommand{\secref}[1]{Section~\ref{sec:#1}}

\begin{document}
\title{Multi-User Millimeter Wave Channel Estimation Using Generalized Block OMP Algorithm}

\author{\IEEEauthorblockN{Manoj A}
\IEEEauthorblockA{Research Scholar\\
Indian Institute of Technology, Madras, India\\
Email: ee14d210@ee.iitm.ac.in}
\and
\IEEEauthorblockN{Arun Pachai Kannu}
\IEEEauthorblockA{Associate Professor \\Indian Institute of Technology, Madras, India\\
Email: arunpachai@ee.iitm.ac.in}
}

\maketitle

\begin{abstract}
In a multi-user millimeter (mm) wave communication system, we consider the problem of estimating the channel response between the 
central node (base station) and each of the user equipments (UE). We propose three different strategies: 1) Each UE estimates its channel separately, 2) Base station estimates all the UEs' channels jointly, and 3) Two stage process with estimation done at both UE and base station. Exploiting the low rank nature of the mm wave channels, we propose a generalized block orthogonal matching pursuit (G-BOMP) framework for channel estimation in all the three strategies. Our simulation results show that, the average beamforming gain of the G-BOMP algorithm is higher than that of the conventional OMP algorithm and other existing works on the multi-user mm wave system.
\end{abstract}


%
\IEEEpeerreviewmaketitle

\section{Introduction}
\par The availability of large spectral bandwidth in the under-utilized millimeter (mm) wave frequency bands makes the mm wave communication system a potential candidate for the 5G cellular technology 
\cite{528506, 7572712, 6042312, 6489099}. Equipped with sophisticated analog-digital hybrid architectures, mm wave systems combat path losses by highly directional beamforming. Proper design of the beamforming precoders and combiners require 
the knowledge of the channel state information. 

The mm wave channel with uniform linear array (ULA) at both transmitter and receiver is modeled as weighted sum of array responses for each path \cite{6717211}. Each path is composed of two spatial frequencies which depend on the angle of departure (AoD) at the transmitter and angle of arrival (AoA) at the receiver. Since the number of paths is small compared to the dimension of the ULA, several compressive sensing based channel estimation schemes were developed for single-user mm wave systems in \cite{7636870, 7439748, 7458188}. 

For multi-user mm wave systems, \cite{7752516} proposed a channel estimation strategy at the base station (BS) using fast iterative shrinkage thresholding algorithm (FISTA). In \cite{7178503}, a training scheme was introduced where user equipments (UE)  estimate the channels using orthogonal matching pursuit (OMP). An asymmetric channel estimation method was proposed in \cite{7763421}, where the UE and base station estimate the channels individually (but details of estimation algorithms are lacking). All these methods assume that the spatial frequencies corresponding to the AoA and AoD of each path fall exactly in the grid points of DFT matrices (integer multiples) of ULA sizes. In practice, the spatial frequencies may not fall exactly in the DFT bins and hence spectral leakage occurs. Since the leakage is concentrated around the spatial frequency, we propose a generalized block OMP (G-BOMP) framework for channel estimation in multi-user mm wave systems. Our main contributions in this paper include: 

\begin{itemize}
\item We consider three different training and estimation strategies: 1) Separate estimation at each UE. 2) Joint estimation at the base station. 3) Two stage process with estimation at both mobiles and base station.
\item We propose a G-BOMP framework which can be employed for all the above three strategies.
\item We show that the proposed G-BOMP framework performs better than the conventional OMP algorithm and other existing training/estimation strategies \cite{7752516,7178503}.
\end{itemize}

\par \textbf{Notations:} 
$(.)^*, (.)^T$ and $(.)^H$ indicate conjugation, transpose and hermitian operations respectively. $\otimes$ denotes Kronecker product and $\vecc(\vec{A})$ gives a vector obtained by vertical concatenation of columns of the matrix $\vec{A}$. $\vec{I}_K$ and $\vec{F}_K$ denote an identity matrix and unitary FFT matrix respectively, of size $K \times K$.

\section{System Model} 

\subsection{Channel Model}
\par Consider a mm wave communication system comprising of a base station and $L$ user equipments. Let the BS be equipped with a ULA consisting of $N_b$ antenna elements and let each UE contain a ULA with $N_u$ antenna elements.  
We adopt the channel model used in \cite{6717211, 7636870, 7439748, 7458188} and define the channel from the BS to the $i^{th}$ UE as,
\begin{align}
\vec{H}_i = \sqrt{\frac{N_uN_b}{K_i}} \sum\limits_{k = 1}^{K_i} \alpha_{i}(k) \vec{a}_{bi}(k) \vec{a}_{ui}(k)^H,
\label{ch_model}
\end{align}
where $K_i$ is the total number of multi-paths, $\alpha_{i}(k)$ is the gain of $k^{th}$ multi-path linking the $i^{th}$ UE and the BS, and $\vec{a}_{bi}(k)$ and $\vec{a}_{ui}(k)$ are the ULA responses at the BS and the $i^{th}$ UE respectively for the $k^{th}$ multi-path. 
We model $\alpha_{i}(k),\, \forall i = 1, 2, ..., L $ and $\forall k = 1, ..., K_i$ as i.i.d. circular Gaussian with variance $\sigma_{\alpha}^2$. The ULA response vectors are given by,
\begin{align}
\vec{a}_{li}(k) = \frac{1}{\sqrt{N_l}} [1 \, e^{j \Omega_{li}(k)} \, ... \, e^{j(N_l-1) \Omega_{li}(k)}]^T, 
\label{resp}
\end{align}
where $l \in \{b,u\}$, $\Omega_{ui}(k) = 2\pi \frac{d}{\lambda} \sin(\phi_{ui}(k))$, $\Omega_{bi}(k) = 2\pi \frac{d}{\lambda} \sin(\phi_{bi}(k))$, $d$ is the spacing between the antenna elements in the ULA, $\lambda$ is the operating carrier wavelength, $\phi_{ui}(k)$ and $\phi_{bi}(k)$ are the angles of departure (AoD) and arrvial (AoA) respectively, for the $k^{th}$ multi-path corresponding to the channel between BS and $i^{th}$ UE, and are uniform in a subset of $[-\pi,\pi]$.

\subsection{Block Sparse Structure}
\par Each path in \eqref{ch_model} has the array responses \eqref{resp}, which are complex exponentials with spatial frequencies $\Omega_{bi}(k)$ and $\Omega_{ui}(k)$. Hence, the channel matrix $\vec{H}_i$ in \eqref{ch_model} is sparse in Fourier domain. To understand the structure, let us define the $2-$D DFT of $\vec{H}_i$ as, 
\begin{align}
\vec{H}_i^{\omega}  = \vec{F}_b^H \vec{H}_i  \vec{F}_u, 
\label{sp_ch_model}
\end{align}
where $\vec{F}_b$ and $\vec{F}_u$ are DFT matrices of size $N_b$ and $N_u$ respectively. Since AoA $\phi_{bi}(k)$ and AoD 
$\phi_{ui}(k)$ in \eqref{resp} are typically uniformly distributed in a subset of $[-\pi,\pi]$, the spatial frequencies will 
not fall exactly in the DFT bins (i.e., will not be an integer multiple of $\frac{2\pi}{N_b}$ or $\frac{2\pi}{N_u}$). Hence, in the DFT domain, we 
encounter spectral leakage, which is concentrated around the exact spatial frequencies. 

\begin{figure}[h]
\center
\includegraphics[scale = 0.4]{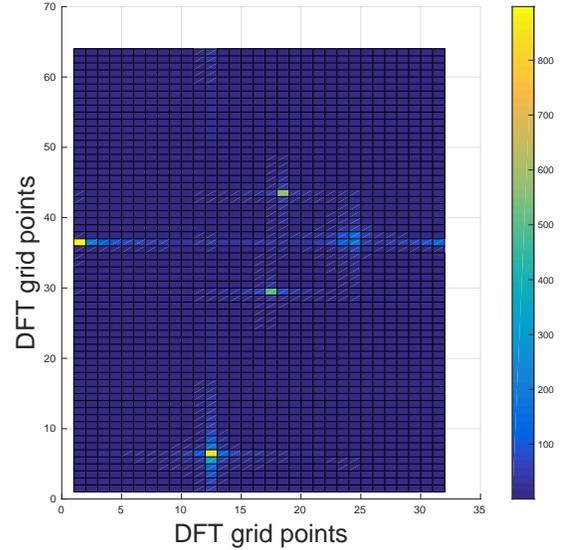}
\caption{Magnitude plot of $\vec{H}_i^{\omega}$ with $N_u = 32$, $N_b = 64$, $K_i = 5$.}
\label{fig:ch_sparse}
\end{figure}

In Figure (\ref{fig:ch_sparse}), 
we illustrate the sparse structure of $\vec{H}_i^{\omega}$, by shading each square based on the sum of magnitude of the DFT grid points which enclose that square. Since the spectral leakage is negligible for the grid points which are far from the actual spatial frequencies,  
$\vec{H}_i^{\omega}$ can be approximated as a $2-$D block sparse matrix, with each path contributing to a non-zero square block, say of size ($b \times b$). Depending on the actual values of the spatial 
frequencies, the non-zero square blocks in $\vec{H}_i^{\omega}$ may or may not be overlapping. 

\subsection{Reception Model}

\par We assume a hybrid beamforming system architecture \cite{6717211} for the BS and for all UEs, with single stream communication. In the downlink, if the BS transmits data symbol $s$, using a beamforming vector $\vec{u}$ (of size $N_b \times 1$), 
then the final output at the $i^{th}$ UE will be,
\begin{align}
z_i = \vec{v}_i^H \vec{H}_i^H \vec{u} s + n_i, \,\, i \in \{1, 2, ..., L\},
\label{BS_to_MS}
\end{align}
where $\vec{v}_i$ (of size $N_u \times 1$) is the beamforming vector used by the $i^{th}$ UE. 
Note that, $\vec{H}_i^H$ is the channel seen from the $i^{th}$ UE to the BS and ${n}_i$ is the circular additive gaussian noise (CWGN) with variance $\sigma_u^2$.

\par Similarly, in the uplink, if $\vec{w}_{ui}$ is the beamforming vector assigned to the ULA at the $i^{th}$ UE, where $i = 1, 2, ..., L$, then the signal observed at the BS, which applies a beamforming weight $\vec{w}_b$, is given by,
\begin{align}
y = \vec{w}_b^H \sum\limits_{i = 1}^L \vec{H}_i \vec{w}_{ui} x_i + n,
\label{MS_to_BS}
\end{align}
where $\vec{H}_i$ is the millimeter wave channel from the BS to the $i^{th}$ UE, $x_i$ is the data symbol sent by the $i^{th}$ UE and ${n}$ is CWGN with variance $\sigma_b^2$. 

\section{Channel Estimation Strategies}  \label{sec:strat}

\subsection{Method 1: Channel Estimation at individual UEs} \label{sec:met1}
\par In this method, each UE estimates its corresponding channel $\vec{H}_i$ based on the training signals sent by the base station. 
Suppose we consider a training phase of duration $M$, where UE makes measurements of the form \eqref{BS_to_MS}, 
with the $m^{th}$ measurement at the $i^{th}$ UE denoted by,
\begin{align*}
z_i^{(m)} = \vec{v}_i^{(m)H} \vec{H}_i^H \vec{u}^{(m)} s^{(m)} + {n}_i^{(m)},\,\, i = 1, ..., L.
\end{align*}
Here, $\vec{u}^{(m)}$ ($\vec{v}_i^{(m)}$) denote the beamforming weight used by the BS (and $i^{th}$ UE) 
during the $m^{th}$ measurement. We assume that the training symbol $s^{(m)} = 1, \, \forall m \in \{1, 2, ..., M\}$. 
With $\vec{h}_i = \vecc(\vec{H}_i)$, above equation can be re-written as,
\begin{align*}
z_i^{(m)H} & = (\vec{v}_i^{(m)T} \otimes \vec{u}^{(m)H}) \vec{h}_i + {n}_i^{(m)H}.
\end{align*}
From \eqref{sp_ch_model}, we have 
\begin{align}
\vec{h}_i = \Psi \vec{h}_i^{\omega}, \label{vecw}
\end{align} 
with $\Psi = \vec{F}_u^* \otimes \vec{F}_b$ and $\vec{h}_i^{\omega} = \vecc(\vec{H}_i^{\omega})$. 
Collecting the $M$ observations into a vector,  $\vec{z}_i = [z_i^{(1)H} \, ... \, z_i^{(M)H}]^T$, we get,
\begin{align}
\vec{z}_i  = \vec{A}_i \Psi \vec{h}_i^{\omega} + \vec{n}_{i}, \,\, i = 1, 2, ..., L,
\label{CS_meas_model_1}
\end{align}
where $\vec{n}_i = [ {n}_i^{(1)H} \, ... \,{n}_i^{(M)H} ]^T$ and the $m^{th}$ row of $\vec{A}_i$ is $\vec{v}_i^{(m)T} \otimes \vec{u}^{(m)H}$, $m = 1, ..., M$. For the observation model \eqref{CS_meas_model_1}, we present a generalized block OMP framework 
in \secref{gbomp}, to reconstruct $\vec{h}_i^{\omega}$, which is the vectorized version of a 2-D block sparse matrix $\vec{H}_i^{\omega}$. 
Once the UEs estimate their channels, they also compute the optimal precoding weights to maximize the beamforming gain using singular value decomposition, and feedback the beamforming weights to the BS individually. 

\subsection{Method 2: Joint Channel Estimation at BS} \label{sec:met2}

\par In this method, UEs send training signals simultaneously and $M$ measurements are made at the BS. We set the beamforming vectors of all the UEs to be the same during the training phase, i.e., $\vec{w}_{ui} = \vec{w}_u$, for all $i \in \{1, 2, ..., L\}$. Then, the observation model in \eqref{MS_to_BS} becomes,
\begin{align}
y = \vec{w}_b^H \Big( \sum\limits_{i = 1}^L \vec{H}_i x_i \Big) \vec{w}_u + n.
\label{sys_model}
\end{align}
Above equation \eqref{sys_model} can be re-formulated as,
\begin{align*}
y &= (\vec{w}_u^T \otimes \vec{w}_b^H) \sum\limits_{i=1}^L \vecc(\vec{H}_i)x_i + {n} = (\vec{w}_u^T \otimes \vec{w}_b^H) \vec{H} \vec{x} + n,
\end{align*}
where $\vec{H} = [\vec{h}_1 \, \vec{h}_2 \, ... \, \vec{h}_L]$ and $\vec{x} = [x_1 \, ... \, x_L]^T$. 
Further, we get,
\begin{align}
y = (\vec{x}^T \otimes (\vec{w}_u^T \otimes \vec{w}_b^H)) \vec{h} + {n}, \label{meas_model}
\end{align}
where $\vec{h} = \vecc(\vec{H})$. Using \eqref{vecw}, we have $\vec{h} =  \Phi \vec{h}^{\omega}$, where $\vec{h}^{\omega} = [(\vec{h}_1^{\omega})^T \, ... \, (\vec{h}_L^{\omega})^T]^T$ and $\Phi = \vec{I}_L \otimes \Psi$. 
Suppose $M$ measurements of the form \eqref{meas_model} are obtained, i.e., $y^{(m)} = (\vec{x}^{(m)T} \otimes (\vec{w}_{u}^{(m)T} 
\otimes \vec{w}_{b}^{(m)H})) \Phi \vec{h}^{\omega} + {n}^{(m)}$, $m = 1, 2, ..., M$, and $\vec{y} = [y^{(1)} \,...\, y^{(M)}]^T$, then,
\begin{align}
\vec{y} = \vec{B} \Phi \vec{h}^{\omega} + \vec{n},
\label{CS_meas_model_2}
\end{align}
where the $m^{th}$ row of $\vec{B}$ will be $(\vec{x}^{(m)T} \otimes (\vec{w}_{u}^{(m)T} \otimes \vec{w}_{b}^{(m)H}))$ and $\vec{n}= 
[n^{(1)} \, ... \, n^{(M)}]^T$. 
Since $\vec{h}^{\omega}$ in the observation model \eqref{CS_meas_model_2} is concatenation of vectors from $2-$D 
block sparse matrices, we use G-BOMP framework from \secref{gbomp} to recover channels of all the UEs jointly. Once the channels are estimated and the optimal weights are computed, the BS informs all the UEs about their corresponding beamforming vectors via feedback.

\subsection{Method 3: Two Stage Channel Estimation Strategy} \label{sec:met3}

\par In this method, the estimation process is done in two stages. In the first phase, UEs will estimate the channel using $M_1$ pilots sent by the BS (Method 1). In the second phase, the UEs will assign the estimated optimal beamforming vector to the ULA and transmit $M_2$ pilots to the BS. BS will use these pilots and determine its optimal beamforming weights corresponding to each UE. For this method, the total training duration is $M=M_1 +M_2$ and we do not require any feedback mechanism to convey the optimal weights. 

First phase proceeds as per Method 1 from \secref{met1} with $M_1$ measurements. Now, let $\vec{u}_{\text{opt}}^{[i]}$ be the estimated optimal beamforming vector for the $i^{th}$ UE.  In the second phase, BS obtains $M_2$ measurements with all the UEs choosing their optimal 
weights for precoding.  The $m^{th}$ ($m = 1,..., M_2$) measurement at the BS will then be,
\begin{align}
y^{(m)} = \vec{w}_b^{(m)H} \Big( \sum\limits_{i = 1}^{L} \vec{H}_i \vec{u}_{\text{opt}}^{[i]}\, x_i^{(m)} \Big) + n^{(m)}.
\label{Method_3_eq}
\end{align}
In order to discuss the estimation process at the BS, we make an approximation that $\vec{u}_{\text{opt}}^{[i]}$ will be oriented along the path corresponding to the largest gain $\alpha_{i}(k)$ in \eqref{ch_model}. Assuming that $|\alpha_{i}(1)|$ is 
the largest, we approximate that $\vec{u}_{\text{opt}}^{[i]} \approx \vec{a}_{ui}(1)$ (Note that we have $\vec{u}_{\text{opt}}^{[i]} = \vec{a}_{ui}(1)$ for a single rank channel matrix).  
With this approximation, we get,
\begin{align*}
\vec{H}_i \vec{u}_{\text{opt}}^{[i]}  
\approx \sqrt{\frac{N_uN_b}{K_i}} & \Big[\alpha_{i}(1) \vec{a}_{bi}(1) \\ & + \sum\limits_{k = 2}^{K_i} \alpha_{i}(k) \vec{a}_{bi}(k) \vec{a}_{ui}(k)^H \vec{u}_{\text{opt}}^{[i]} \Big],
\end{align*}
and equation \eqref{Method_3_eq} can be re-formulated as,
\begin{align*}
y^{(m)} \approx \vec{w}_b^{(m)H} \sum\limits_{i = 1}^L \sqrt{\frac{N_u N_b}{K_i}} \alpha_{i}(1) x_i^{(m)} \vec{a}_{bi}(1) + \tilde{n}^{(m)},
\end{align*}
where $\tilde{n}^{(m)} = \vec{w}_b^{(m)H} \Big[ \sum\limits_{i = 1}^L \sum\limits_{m = 2}^{K_i} \alpha_{i}(m) \vec{a}_{bi}(m) \vec{a}_{ui}(m)^{H} \vec{u}_{\text{opt}}^{[i]} \\ x_i^{(m)} \Big] + n^{(m)}$. In the above equation, each $\vec{a}_{bi}(1)$ 
is a complex exponential with spatial frequency $\Omega_{bi}^{(1)}$. 
Denoting $\sqrt{\frac{N_uN_b}{K_i}} \alpha_{i}(1) \vec{a}_{bi}(1)$ as $\vec{c}_i$, we get,
\begin{align*}
y^{(m)} &= \vec{w}_b^{(m)H} [\vec{c}_1 \, \vec{c}_2 \, ... \, \vec{c}_L] \vec{x}^{(m)} + \tilde{n}^{(m)}, 
\end{align*}
where $\vec{x}^{(m)} = [x_1^{(m)} \, x_2^{(m)} \, ... \, x_L^{(m)}]^T$. 
Suppose $\vec{c}_i = \vec{F}_b \vec{c}_i^{\omega}$, where $\vec{c}_i^{\omega}$ is a $1-$D block sparse vector with spectral leakage concentrated around
the frequency $\Omega_{bi}^{(1)}$ and $\vec{y} = [y^{(1)} \, ... \, y^{(M_2)}]^T$, then,
\begin{align}
\vec{y} = \vec{D} \Gamma \vec{c}^{\omega} + \tilde{\vec{n}},
\label{Method_3_CS_model}
\end{align}
where the $m^{th}$ row of $\vec{D}$ is $(\vec{x}^{(m)T} \otimes \vec{w}_b^{(m)H}), \, m = 1, ..., M_2$, $\tilde{\vec{n}} = [\tilde{n}^{(1)} \, ... \, \tilde{n}^{(M_2)}]^T$, $\Gamma = \vec{I}_L \otimes \vec{F}_b$ and $\vec{c}^{\omega} = [(\vec{c}_1^{\omega})^T \, ... \, (\vec{c}_L^{\omega})^T]^T$. Observation model \eqref{Method_3_CS_model} can again be solved using generalized block OMP framework discussed in \secref{gbomp} by specializing it to the $1-$D case. Here, we directly obtain the optimal beamforming weights $\vec{a}_{bi}(1)$ 
without estimating the entire channel at the BS.  Note that effective noise term in \eqref{Method_3_CS_model} includes contributions from all multi-path components except the strongest one. This multi-path interference will become the limiting factor when noise power $\sigma_u^2$ is small in the
estimation model in \eqref{Method_3_CS_model}.

\subsection{Training Beamforming vectors and Signals}

All the three estimation strategies involve certain beamforming weights (at each measurement $m$) 
to be used at the BS such as $\{\vec{w}_b^{(m)}, \vec{u}^{(m)} \}$ and at the UE such as $\{\vec{w}_u^{(m)}, \vec{v}_i^{(m)}\}$ and 
training symbols used at the UE $\{x_i^{(m)}\}$. We generate all the entries in the beamforming vectors and training symbols using i.i.d. 
Bernoulli distribution with entries being $\{\pm 1\}$ equally likely. We then normalize the beamforming vectors to have unit norm. The training beamforming vectors and training symbols are revealed to both BS and UE so that they can perform the channel estimation. Since our training beamforming vectors are binary, they can be (very) easily implemented in RF chains when compared with other training methods which use complex phases (of the form $e^{j\theta}$) 
as in \cite{7178503} or \cite{7752516}.

\section{Generalized Block OMP framework} \label{sec:gbomp}
\par 
We consider the block sparse signal recovery framework for the estimation of channel matrices $\vec{H}_i^w$ from our measurement models. 
However, the block OMP algorithm from \cite{Eldar:TSP:10} is developed for the case when the block sparse vector is \emph{apriori} partitioned into 
\emph{disjoint} sub-blocks, out of which few are non-zero. In our model, such disjoint apriori partitioning is not possible since the non-zero blocks (squares) in $\vec{H}_i^w$ depend on the actual spatial frequencies of each path. Hence, we propose a generalized version of block OMP algorithm (G-BOMP), which will be applied to solve the mm wave channel estimation problem, for all the three strategies in \secref{strat}. 

\par Consider a $g_1 \times g_2$ matrix, $\vec{G} = \vec{S}_1 \vec{E} \vec{S}_2$, where $\vec{E}$ (of size $P \times Q$) is a $K'-$block sparse matrix, i.e., there exists at most $K'$ non-zero blocks of order $b \times b$ at arbitrary locations,  with all remaining entries being zero. 
Consider the noisy observation model $\vec{y} = \vec{A} \vec{S} \vec{e} + \vec{n}$, where $\vec{S} = (\vec{S}_2^T \otimes \vec{S}_1)$ and $ \vec{e} = \vecc(\vec{E})$ and $\vec{A}$ is the measurement matrix of size $m' \times PQ$. 

\par First we define sub-blocks of the matrix $\vec{E}$. The illustration is given below for $b=2$. Define sets $\mathcal{P} = \{1, ..., P\}$ and $\mathcal{Q} = \{1, ..., Q\}$. Let $\vec{B}_{p,q}$ with $p \in \mathcal{P}$, $q \in \mathcal{Q}$, define a sub-block of size $b \times b$ with top left entry being $e_{p,q}$. For example, in the below picture $\vec{B}_{1,2}$ is a square containing the entries $\{e_{1,2}, e_{1,3}, e_{2,2}, e_{2,3} \}$ and  $\vec{B}_{P,4}$ contains the entries $\{e_{P,4}, e_{P,5}, e_{1,4}, e_{1,5} \}$. Let the index set $\mathcal{J}_{p,q}$ denote the locations of the entries of $\vec{B}_{p,q}$ in the vector $\vec{e}$. From the illustration below, $\mathcal{J}_{1,1} = \{1,2,P+1,P+2\}$. For a given $p$ and $q$, $\mathcal{J}_{p,q}$ is obtained as follows: Let $r_{k_1,k_2} = p + k_1 + (k_2 + q - 1)P$, where $k_1, k_2 \in \{0, 1, ..., b-1\}$. Suppose $r_{k_1,k_2}$ exceeds $(q + k_2)P$, then modify $r_{k_1,k_2}$ as $ r_{k_1,k_2} - P$. Refine $r_{k_1,k_2} = r_{k_1,k_2} \,\text{mod}\, PQ$. Then, $\mathcal{J}_{p,q} = \bigcup\limits_{k_1=0}^{b-1} \bigcup\limits_{k_2=0}^{b-1} \{r_{k_1,k_2}\}$. 

\begin{center}
\begin{tikzpicture}
      \matrix [matrix of math nodes,left delimiter=(,right delimiter=)] (m)
       {
            e_{1,1} &e_{1,2} &e_{1,3} &e_{1,4} & e_{1,5} &  ... & e_{1,Q}\\
            e_{2,1} &e_{2,2} &e_{2,3} &e_{2,4} & e_{2,5} &  ...& e_{2,Q}\\
            ... & ... & ... & ... & ... & ... & ... \\
            e_{P,1} &e_{P,2} &e_{P,3} &e_{P,4} & e_{P,5} &  ...& e_{P,Q}\\
        };  
        \draw[color=green,thick] (m-1-1.north west) -- (m-1-2.north east) -- (m-2-2.south east) -- (m-2-1.south west) -- (m-1-1.north west);
        \draw[color=blue,thick] (m-1-2.north west) -- (m-1-3.north east) -- (m-2-3.south east) -- (m-2-2.south west) -- (m-1-2.north west);
        \draw[color=yellow,thick] (m-1-3.north west) -- (m-1-4.north east) -- (m-2-4.south east) -- (m-2-3.south west) -- (m-1-3.north west);
        \draw[color=red,thick] (m-1-7.north west) -- (m-1-7.south west) -- (m-1-7.south east);
        \draw[color=red,thick] (m-1-1.north east) -- (m-1-1.south east) -- (m-1-1.south west);
        \draw[color=red,thick] (m-4-1.north west) -- (m-4-1.north east) -- (m-4-1.south east);
        \draw[color=red,thick] (m-4-7.south west) -- (m-4-7.north west) -- (m-4-7.north east);
        \draw[color=black] (m-4-4.south west)--(m-4-4.north west) -- (m-4-5.north east) --(m-4-5.south east);
        \draw[color=black] (m-1-4.north west)--(m-1-4.south west) -- (m-1-5.south east) --(m-1-5.north east);
    \end{tikzpicture}
\end{center}

Let $\mathcal{B} = \{\vec{B}_{p,q}\}_{p \in \mathcal{P}, q \in \mathcal{Q}}$ be the collection of valid sub-blocks of $\vec{E}$, that is, the set of all the possible non-zero blocks in $\vec{E}$. Let $\mathcal{J} = \{\mathcal{J}_{p,q}\}_{p \in \mathcal{P}, q \in \mathcal{Q}}$ denote the corresponding collection of index sets.

The inputs to the G-BOMP algorithm are $\vec{y}$, $\bar{\vec{A}} = \vec{A} \vec{S}$, a collection of valid sub-blocks $\mathcal{B}$ and the corresponding index sets $\mathcal{J}$ and a stopping criterion. 

\begin{enumerate}
\item Initialize variables: $\vec{r} = \vec{y}$, $\vec{M} = \vec{0}_{P \times Q}$ (all zero matrix) 
and $\mathcal{I} = \emptyset$. Iteration $t =1$; 
\item Compute: $\vec{b} = \bar{\vec{A}}^H \vec{r}$.
\item Assign $[\vec{M}]_{p,q}$ as $||\vec{b}_{\mathcal{J}_{p,q}}||_2$, for every $\mathcal{J}_{p,q} \in \mathcal{J}$. Here, $\vec{b}_{\mathcal{J}_{p,q}}$ is a sub-vector containing entries from $\vec{b}$ located at positions dictated by the index set $\mathcal{J}_{p,q}$.
\item Evaluate $(\lambda_r(t), \lambda_c(t)) = \arg \max\limits_{(p,q) \in \mathcal{P} \times \mathcal{Q}} [\vec{M}]_{p,q}$.
\item Store the identified index points: $\mathcal{I} = \mathcal{I} \cup \mathcal{J}_{\lambda_r(t),\lambda_c(t)}$. 
\item Compute $\vec{x}_t = \arg \min\limits_{\vec{x}} ||\vec{y} - \bar{\vec{A}}_{\mathcal{I}} \vec{x}||_2$, where $\bar{\vec{A}}_{\mathcal{I}}$ is a sub-matrix of $\bar{\vec{A}}$, containing those columns of $\bar{\vec{A}}$ indexed by $\mathcal{I}$.
\item Update the residue as, $\vec{r} = \vec{y} - \bar{\vec{A}}_{\mathcal{I}} \vec{x}_t$.
\item Increment $t$ by $1$. If the stopping criterion described below is satisfied, then stop. Else, go to step $2$.
\end{enumerate}
When the algorithm stops, the estimate of $\vec{e}$ is obtained as, $\hat{\vec{e}} = (\bar{\vec{A}}_{\mathcal{I}}^H \bar{\vec{A}}_{\mathcal{I}})^{-1} \bar{\vec{A}}_{\mathcal{I}}^H \vec{y}$. The stopping criterion is: $t \leq K'$ or $||\bar{\vec{A}}^H \vec{r}||_{\infty}^2 \leq \tau$, where the threshold $\tau$ is appropriately chosen based on the operating SNR and the number of measurements used.
\par In a typical mm wave communication system, since the number of multi-paths is very small \cite{6732923, 7417864}, we assume that $K_i \leq K_{\max}$, for all $i = 1, ..., L$ and for some positive integer $K_{\max}$.  
\par The channel estimation models in \secref{strat} can be formulated in the G-BOMP framework directly. For instance, when the individual UEs estimate their corresponding channels (in Method 1 \secref{met1} and first phase in Method 3 in \secref{met3}), we have $\vec{E} = \vec{H}_i^{\omega}$ and $K' = K_{\max}$. In the case of BS jointly estimating all the channels (Method 2 in \secref{met2}), we have $\vec{E} = [\vec{H}_1^{\omega},\cdots,\vec{H}_L^{\omega}]$ and $K' = LK_{\max}$. The $1-D$ version of G-BOMP can be easily obtained by setting $Q=1$, which is needed for the second phase of Method 3. In all these cases, the set of all the possible non-zero blocks can be specified by 
considering the range of spatial frequencies in the channel model (and by neglecting the leakage outside $b \times b$ squares centered around the spatial frequencies). 

\begin{figure}[h]
\center
\includegraphics[width = 8cm, height = 8cm]{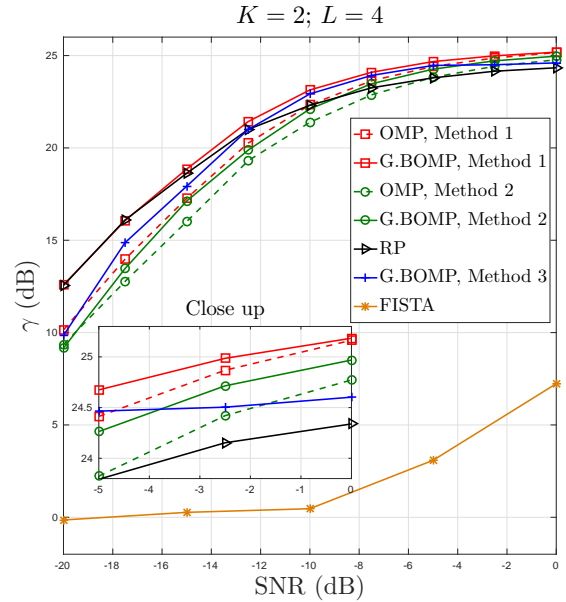}
\caption{$\gamma$ (in dB) Vs SNR (in dB), given $M = 225$, $K = 2$ and $L = 4$.}
\label{fig:Avg_gain_vs_snr_L2_L4}
\end{figure}

\section{Simulation Results}
\par We studied the performance of our G-BOMP algorithm via simulations. The parameters considered are: $N_b = 32 = 2N_u$, $\frac{d}{\lambda} = \frac{1}{2}$, $b = 2$, block size for $1-$D G-BOMP algorithm is $4$, $\phi_{bi}(k), \phi_{ui}(k)$ are i.i.d. random variables uniform in $[-\frac{\pi}{2},\frac{\pi}{2}]$, $\forall k = 1, 2, ..., K_i$ and $i = 1, ..., L$, $\sigma_{\alpha}^2 = 1$. We assume $K_i = K,\, \forall i = 1, 2, ..., L$, $K_{\max} = 3$ and $\sigma_u^2 = \sigma_b^2 = \sigma^2$. Further, in our simulations, we assume the mm wave channels, $\vec{H}_i,\,\forall i = 1,...,L$, to be, $\vec{H}_i = \vec{F}_{bb}\vec{H}_i^{\omega} \vec{F}_{uu}^H$, where $\vec{F}_{bb}$ ($\vec{F}_{uu}$) is an $N_b \times 2N_b$ ($N_u \times 2N_u$) Fourier matrix. We studied the performance of our G-BOMP algorithm in terms of the average beamforming gain ($\gamma$) achieved, which is defined as,
\begin{align}
\gamma = \frac{1}{L} \sum\limits_{i = 1}^L |\hat{\vec{w}}_{\text{opt}}^{(i)H}\vec{H}_i \hat{\vec{f}}_{\text{opt}}^{(i)}|^2,
\end{align}
where $\hat{\vec{w}}_{\text{opt}}^{(i)}$ and $\hat{\vec{f}}_{\text{opt}}^{(i)}$ are the left and the right singular vectors of $\hat{\vec{H}}_i$ (the estimate of $\vec{H}_i$). We compare the performance of our G-BOMP algorithm with the OMP algorithm, the FISTA method \cite{7752516} and a training scheme proposed in \cite{7178503} (Random Phase (RP) method).

\begin{figure}[h]
\center
\includegraphics[width = 8cm, height = 6cm]{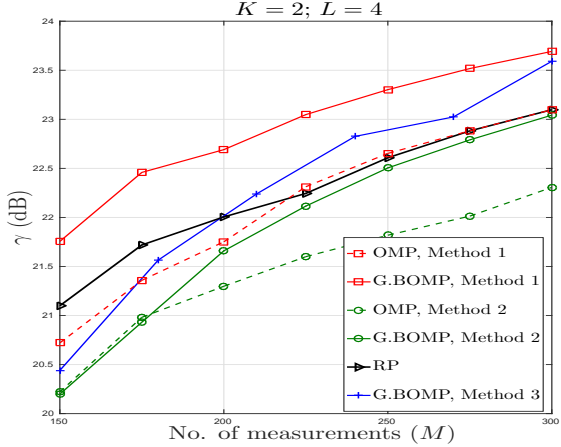}
\caption{$\gamma$ plotted as a function of $M$, for $K = 2$, $L = 4$ and SNR = $-10$dB.}
\label{fig:Avg_gain_vs_meas_L2_L4}
\end{figure}

\par Figure (\ref{fig:Avg_gain_vs_snr_L2_L4}) plots $\gamma$ in (dB) as a function of SNR, which is defined as $\frac{1}{\sigma^2}$. Parameters used are: $L = 4$, $K = 2$, $M = 225$, $M_1 = 125$ and $M_2 = 100$. We observe the following:
\begin{enumerate}
\item $\gamma$ increases with SNR for all schemes and G-BOMP is better than other techniques in each method. However, the rate at which $\gamma$ increases, decays with SNR.
\item $\gamma$ of Method 1 is higher than that achieved in Methods 2 and 3, reason being that each UE estimates its channel without any interference from the other UEs. Also, since Method 2 jointly re-constructs the channels, i.e., the entity to be estimated is of larger dimension, it requires larger $M$ to achieve the same level of performance as that of Method 1.
\item In the low SNR regime, Method 3 is better than Method 2. But Method 2 becomes superior to Method 3 at higher values of SNR, e.g., SNR $\geq -3$dB. This is because when SNR is high, the direction with the largest gain sees increased interference from the paths corresponding to other spatial frequencies in Method 3, as per our remarks following equation \eqref{Method_3_CS_model}.
\end{enumerate} 
\par Figure (\ref{fig:Avg_gain_vs_meas_L2_L4}) analyzes the variaion of $\gamma$ w.r.t $M$. We assume $M_1 = \frac{2}{3}M$, $K = 2$, $L = 4$ and SNR = $-10$dB. As expected, the average beamforming gain of all methods increases with the value of $M$. In particular, $\gamma$ of G-BOMP in a method is found to exceed $\gamma$ of other schemes for that method.

\section{Conclusion}
\par In this article we presented three different strategies of estimating mm wave channels in a multi-user scenario, namely: 1) UEs estimating their channels separately, 2) BS jointly estimating all the channels, and 3) Two stage process where both UEs and the BS estimate the channels. We exploited the sparse nature of mm wave channels and proposed a generalized BOMP algorithm to estimate them in all the three strategies. Our simulation results show that our G-BOMP algorithm performs better compared to the OMP algorithm and other prior works in terms of the average beamforming gain achieved.

\bibliographystyle{ieeetr}
\bibliography{beamforming_ref.bib}

\end{document}